\pdfoutput=1

\documentclass[11pt]{article}

\usepackage[preprint]{acl}

\usepackage{times}
\usepackage{latexsym}
\usepackage{paralist}
\usepackage{amsmath}
\usepackage{amssymb}
\usepackage{enumitem}
\usepackage[normalem]{ulem}

\usepackage[T1]{fontenc}

\usepackage[utf8]{inputenc}

\usepackage{microtype}

\usepackage{inconsolata}

\usepackage{graphicx}

\usepackage[ruled, vlined, linesnumbered]{algorithm2e}
\usepackage{bm}
\usepackage{bbm}
\usepackage{amsthm}
\usepackage{amsmath}
\usepackage{mathrsfs} 
\usepackage{tabularx}
\usepackage{graphicx}
\usepackage{multirow}
\usepackage{xspace}
\usepackage{xcolor}
\usepackage{enumitem}
\usepackage{placeins}
\usepackage{soul}
\usepackage{mathtools}
\usepackage{booktabs}
\usepackage{hyperref}
\usepackage{subcaption}
\usepackage{pgfplots}
\usepackage{csquotes}
\usepackage{cleveref}
\usepackage{balance}
\usepackage{multicol}
\usepackage{makecell}
\usepackage{amssymb}
\usepackage[switch]{lineno}
\DeclareUnicodeCharacter{2212}{−}
\usepgfplotslibrary{groupplots,dateplot}
\usetikzlibrary{patterns,shapes.arrows}
\pgfplotsset{compat=newest}

\newcommand{\ours}{{fMRLRec}\xspace}

\title{
Train Once, Deploy Anywhere: Matryoshka Representation Learning for Multimodal Recommendation
}

\author{Yueqi Wang\textsuperscript{1}\thanks{Both authors contributed equally to this research.} \quad Zhenrui Yue\textsuperscript{2}\footnotemark[1] \quad Huimin Zeng\textsuperscript{2} \quad Dong Wang\textsuperscript{2}\thanks{Corresponding Author} \quad Julian McAuley\textsuperscript{3} \\
\textsuperscript{1}University of California, Berkeley \quad \textsuperscript{2}University of Illinois Urbana-Champaign \\
\textsuperscript{3}University of California, San Diego \\
\texttt{yueqi@berkeley.edu} \quad \texttt{\{zhenrui3, huiminz3, dwang24\}@illinois.edu} \\
\texttt{jmcauley@ucsd.edu} \\}

\begin{document}
\maketitle
\begin{abstract}
Despite recent advancements in language and vision modeling, integrating rich multimodal knowledge into recommender systems continues to pose significant challenges. This is primarily due to the need for 
efficient recommendation, which requires adaptive and interactive responses. 
In this study, we focus on sequential recommendation and introduce a lightweight framework called full-scale Matryoshka representation learning for multimodal recommendation (\ours). Our \ours captures item features at different granularities, learning informative representations for efficient recommendation across multiple dimensions. To integrate item features from diverse modalities, \ours employs a simple mapping to project multimodal item features into an aligned feature space. Additionally, we design an efficient linear transformation that embeds smaller features into larger ones, substantially reducing memory requirements for large-scale training on recommendation data. Combined with improved state space modeling techniques, \ours scales to different dimensions and only requires one-time training to produce multiple models tailored to various granularities. We demonstrate the effectiveness and efficiency of \ours on multiple benchmark datasets, which consistently achieves superior performance over state-of-the-art baseline methods. We make our code and data publicly available at https://github.com/yueqirex/fMRLRec.
\end{abstract}
\section{Introduction}
Recent advancements in language and multimodal modeling demonstrates significant potential for improving recommender systems~\cite{touvron2023llama2, liu2023visual, OpenAI2023GPT4TR, reid2024gemini}. Such progress can be largely attributed to: (1)~language~/~vision features can provide additional descriptive information for understanding user preference and item characteristics (e.g. item descriptions); and (2)~generic language capabilities acquired through language and vision pretraining tasks could be transferred for use in recommendation systems. Consequently, language and multimodal representations provide a robust foundation for enhancing the contextual relevance and accuracy of recommendations~\cite{li2023text, geng2023vip5, yue2023llamarec, wei2024llmrec}.

Despite performance improvements, different recommendation scenarios (e.g., centralized or federated recommender systems) often require varying granularities (i.e., model~/~dimension sizes) in item representations to achieve the balance between performance and efficiency~\cite{han2021deeprec, luo2022personalized, xia2023efficient, zeng2024federated}. For instance, larger dimensions are typically required to encode language and vision features for fine-grained understanding and generation tasks, although marginally lower performance can often be achieved using considerably smaller feature sizes~\cite{kusupati2022matryoshka}.
To identify the optimal granularity for specific use cases in recommendation systems, methods like grid search or adaptive search heuristics are frequently utilized in training~\cite{wang2024auto}. However, such searches can lead to substantial training expenses or fail to identify the optimal model, particularly when given a large configuration space and constrained computational resources. Therefore, a train-once and deploy-anywhere solution is optimal for the efficient training of recommender systems, which should ideally meet the following criteria:
\begin{enumerate}
    \item Training is only need once to yield multiple models of different sizes corresponding to various performance and memory requirements;
    \item Training and inference should demand no more computational costs than training a single large model, allowing deployment of various model sizes at inference time.
\end{enumerate} 

Inspired by Matryoshka Representation Learning (MRL)~\cite{kusupati2022matryoshka}, we introduce a lightweight multimodal recommendation framework named full-scale Matryoshka Representation Learning for Recommendation (\ours). \ours embeds smaller vector/matrix representations in larger ones like Matryoshka dolls and is only trained once without additional computation costs. Different from MRL that only embeds smaller final-layer activations into larger ones during training, \ours pushes the efficiency of MRL training by introducing an efficient linear transformation that embeds both smaller weights and activations into larger ones, thereby reducing memory costs associated with both aspects. This approach is particularly effective for training recommender systems on large-scale data, offering a highly efficient framework for multi-granularity model training. Combined with further improvements in state-space modeling represented by~\cite{yue2024linear, orvieto2023resurrecting, gu2023mamba}, the linear recurrence architecture in \ours delivers both effectiveness and efficiency in recommendation performance across various benchmark datasets. We summarize our contributions below\footnote{We adopt publicly available datasets in our experiments and will release our implementation upon publication.}:
\begin{enumerate}
    \item We introduce a novel training framework for multimodal sequential recommendation (\ours), which provides an efficient paradigm to learn models of varying granularities within a single training session.
    \item \ours introduces an efficient linear transformation that reduces memory costs by embedding smaller features into larger ones. Combined with improved state-space modeling, \ours achieves both efficiency and effectiveness in multimodal recommendation.
    \item We show the effectiveness and efficiency of our \ours on benchmark datasets, where the proposed \ours consistently outperforms state-of-the-art baselines with considerable improvements in training efficiency and recommendation performance.
\end{enumerate}
\section{Related Works}

\subsection{Multimodal Recommendation}
Language and multimodal models are applied as recommender systems to understand user preferences and item properties~\cite{hou2022towards, li2023text, he2016vbpr, wei2023multi}. Current language-based approaches leverage pretrained models to improve item representations or re-rank retrieved items~\cite{chen2023palr, li2023gpt4rec, luo2023recranker, yue2023llamarec, xu2024prompting}. For example, VQ-Rec utilizes a language encoder and vector quantization to improve item features in cross-domain recommendation~\cite{hou2023learning}. To further incorporate visual data, existing methods focus on developing strategies that extracts informative user~/~item representations~\cite{wei2019mmgcn, tao2020mgat, wang2023missrec, wei2024towards, wei2024llmrec}. For instance, VIP5 leverages a pretrained transformer with additional vision encoder to learn user transition patters and improve recommendation performance~\cite{geng2023vip5}. However, current models are not tailored to accommodate flexible item attributes or modalities, nor are they optimized for scalable model sizes and efficient inference. Moreover, multimodal approaches require substantial computational resources and separate training sessions for each model, rendering them largely ineffective for real-world applications. To address this, we introduce a lightweight multimodal recommendation framework in \ours, offering multiple model sizes within a single training session and efficient inference capabilities across various scenarios.

\subsection{Matryoshka Representation Learning}
Matryoshka representation learning (MRL) constructs embeddings at different granularities using an identical model, thereby providing adaptability to varying computational resources without additional training~\cite{kusupati2022matryoshka}. MRL proposes nested optimization of vectors in multiple dimensions using shared model parameters, demonstrating promising results on multiple downstream tasks and further applications~\cite{cai2024matryoshka, hu2024matryoshka, li20242d}. Nevertheless, training MRL models demands additional memory for activations in its nested optimization, posing challenges for training recommender systems with large batches on extensive data. Furthermore, MRL remains unexplored for sequential modeling and efficient multimodal recommendation. As such, our \ours aims to provide an adaptive framework for learning recommender systems using arbitrary modalities, delivering both efficacy and efficiency in multimodal sequential recommendation.
\section{Methodologies}
\label{sec:method}
\subsection{Problem Statement}
We present \ours with a research focus in multimodal sequential recommendation. Formally, given a user set $\mathcal{U}=\{u1, u2,...,u_{|U|}\}$ and an item set $\mathcal{V}=\{v1, v2,...,v_{|V|}\}$, user $u$'s interacted item sequence in chronological order is denoted with $\mathcal{S}_u=[v_1^{(u)}, v_2^{(u)},...,v_n^{(u)}]$, where $n$ is the sequence length. The sequential recommendation task is to predict the next item $v_{n+1}^{(u)}$ that user $u$ will interact with. Mathematically, our objective can be formulated as the maximization of the probability of the next interacted item $v_{n+1}^{(u)}$ given $\mathcal{S}_u$:
\begin{equation}
 p(v_{n+1}^{(u)}=v|\mathcal{S}_u)
\end{equation}

\subsection{Full-Scale Matryoshka Representation Learning for Recommendation}
\label{sec:fMRLRec}
\begin{figure}[t]
    \centering
    \includegraphics[width=\linewidth]{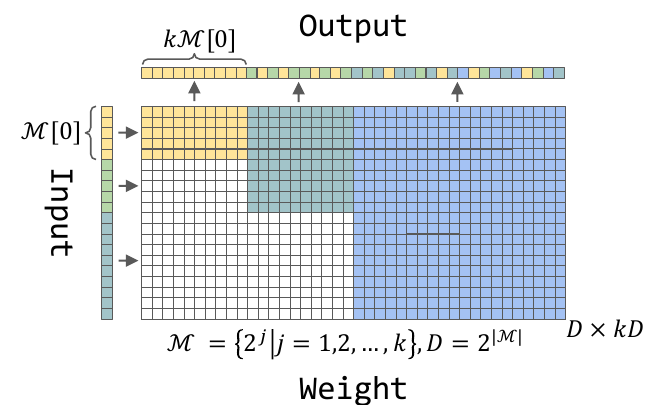}
    \caption{fMRLRec-based weight design, white cells indicate zeros and arrows show vector-matrix multiplication. Input slice $[0:m]$ is only relevant to weight matrix slice $[0:m,0:km]$ during training, convenient for variously-sized model weights extraction during inference time.} 
    \label{fig:fMRL weight}
\end{figure}

In this section, we elaborate on how we design the full-scale Matryoshka Representation Learning for multimodal sequential recommendation (\ours). The majority of model parameters in neural networks can be represented with a set of 2-dimensional weights $\mathcal{W}=\{W_1, W_2, \ldots, W_n\}$ where $ W_i \in \mathbb{R}^{d_1 \times d_2}, i \in \{1, 2, \ldots, n\}$, regardless of specific model architecture. Intuitively, \ours aims to design the $W_i \in \mathcal{W}$ s.t. models of differently sizes $\mathcal{M}=[2, 4, 8, 16, \ldots, D]$ are trained only once at the same cost of only training size-$D$ model. After training, any model sizes in $\mathcal{M}$ can be extracted from the size-$D$ model to form \textit{independent} small models for deployment. To achieve this goal, \ours allows small models to be \textit{embedded} in the largest model. Define sequential input as $X_i \in \mathbb{R}^{B \times L \times D}$ to be processed by $\mathcal{W}$, where $B$ is batch size, $L$ is item sequence length and $D$ is the embedding size, there are three cases for the shape of $W_i \in \mathbb{R}^{d_1 \times d_2}$, denoted as $D(W_i)$,
\begin{equation}
\label{dim cases}
\mathbf{D}(\mathbf{W}_i) = \begin{cases}
    \mathbf{D} \times \mathbf{k}\mathbf{D} & \text{if } \mathbf{d}_1 < \mathbf{d}_2 \\
    \mathbf{k}\mathbf{D} \times \mathbf{D} & \text{if } \mathbf{d}_1 > \mathbf{d}_2 \\
    \mathbf{D} \times \mathbf{D} & \text{if } \mathbf{d}_1 = \mathbf{d}_2
\end{cases}
\end{equation}
Here, we assume $k \in \mathbb{Z}^+/\{1\}$ to ease the derivation since $W_i$ often indicates linear up/down scaling by an integer $k$ times (e.g., post-attention MLPs in transformer).

For case 1 where $D(W_i)=D \times kD$ and $X_i \in \mathbb{R}^{B \times L \times D}$, $X_iW_i$ indicates an up scale. We define the $j$'s slice of $X_i$ as $\mathbf{X}_i^{(j)} = \mathbf{X}_i[0:\mathbf{M}[j]]$ and the $j$'s slice of $W_i$ as

\[
\mathbf{W}_i^{(j)} = 
\begin{cases}
    \mathbf{W}_i[0:\mathbf{M}[0], 0:\mathbf{k}\mathbf{M}[0]] & \quad \text{if } j=0 \\
    \mathbf{W}_i[0:\mathbf{M}[j], \mathbf{k}\mathbf{M}[j-1] & \quad \text{if } j>0 \\
    \quad\quad\;\; :\mathbf{k}\mathbf{M}[j]]
\end{cases}
\]

For case 2 where $D(w_i)=kD \times D$ and the corresponding input $X_i \in \mathbb{R}^{B \times L \times kD}$, $X_iW_i$ indicates a down scale. We define the $j$'s slice of sequential input $X_i$ as $\mathbf{X}_i^{(j)} = \mathbf{X}_i[0:2\mathbf{M}[j]]$ and the $j$'s slice of $W_i$ as

\[
\mathbf{W}_i^{(j)} = \begin{cases}
    \mathbf{W}_i[0:\mathbf{k}\mathbf{M}[0], 0:\mathbf{M}[0]] & \text{if } j = 0 \\
    \mathbf{W}_i[0:\mathbf{k}\mathbf{M}[j], \mathbf{M}[j-1] & \text{if } j > 0 \\
    \quad\quad\;\; :\mathbf{M}[j]]
\end{cases}
\]

For case3 where $D(w_i)=D \times D$, assign $k=1$ for any of above two cases yields $\mathbf{W}_i^{(j)}$.

Then, we perform matrix multiplication between $X_i^{(j)}$ and $W_i^{(j)}$ followed by concatenation along dimension $j$ to form the output
\begin{equation}
\label{compute yi}
\mathbf{Y}_i = [\mathbf{X}_i^{(0)}\mathbf{W}_i^{(0)}, \ldots, \mathbf{X}_i^{(z)}\mathbf{W}_i^{(z)}]
\end{equation}
where $z=\log(D/2)$. Refer to figure \ref{fig:fMRL weight} for case 1 of this process.
\begin{figure*}[t]
  \centering
  \includegraphics[width=0.9\textwidth]{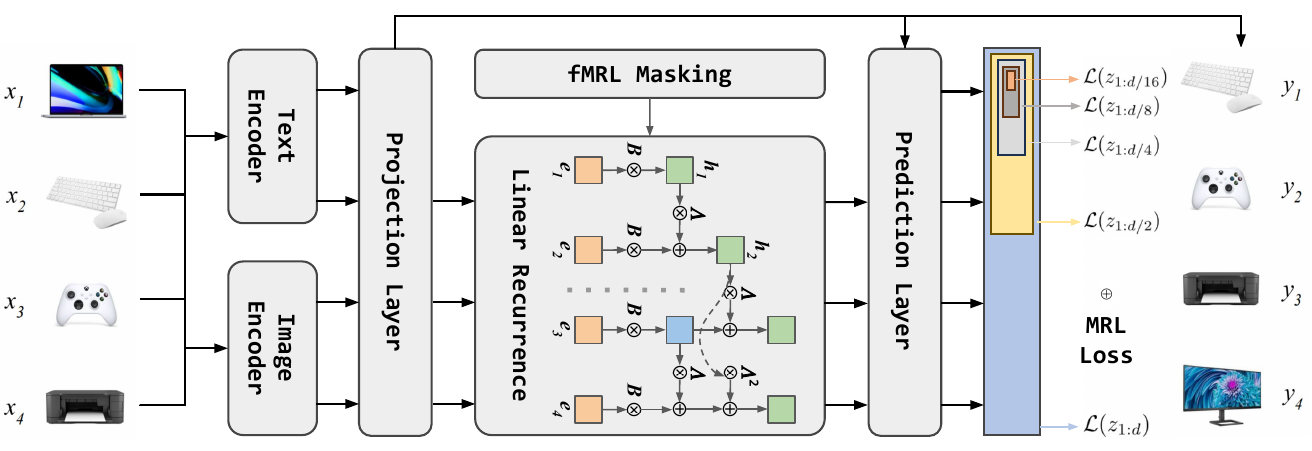}
  \caption{The overall architecture for \ours.}
  \label{fig:main_structure}
\end{figure*}
\paragraph{The \ours Operator} Instead of computing equation \ref{compute yi}, we would like the chunk/slice-wise multiplication of $X_i^{(j)}W_i^{(j)}$ for all $j={1,2,\ldots,\log(D/2)}$ is computed by \textit{one} forward pass to derive output $Y_i$. Specifically, we create a padding mask $P_i(\mathcal{M})$ of the same size as $W_i$ that 
\begin{equation}
\mathbf{P}_i(\mathcal{M}) = \{p_{rs}=0 | w_{rs} \in \mathbf{W}_i, w_{rs} \notin \mathbf{W}_i^{(j)}\}
\end{equation}
Then we define the \ours operator as:
\begin{equation}
\text{\ours}(\mathbf{W}_i, \mathcal{M})=\mathbf{P}_i(\mathcal{M}) \odot \mathbf{W}_i
\end{equation}
Thus, $X_i \cdot \text{\ours}(\mathbf{W}_i, P_i(\mathcal{M}))$ is equivalent to perform equation \ref{compute yi} but only with one time multiplication of $X_i$ and masked $W_i$. See figure \ref{fig:fMRL weight} for an illustration of the \ours operator.

In summary, given a neural network represented by $\mathcal{W}=\{W_1, W_2, \ldots W_n\}$ where $W_i \in \mathbb{R}^{d1 \times d2}$ and a set of sizes $\mathcal{M}=\{2,4,8, \ldots, D\}$, we could find an \ours-slicing of $\mathcal{W}$ such that the first $\mathcal{M}[j]$ elements of input $X_i$ is only processed by corresponding chunks in $W_i$. After the model is trained, we take the first $[0:\mathcal{M}[j],0:k \mathcal{M}[j]$ or $[0:k \mathcal{M}[j],0:\mathcal{M}[j]]$ (depending on the cases in equation \ref{dim cases}) slice for each $W_i$ to form \textit{independent} small models called \ours-series models for inference. Also refer to the upper left of figure \ref{fig:fMRL weight} for the slicing process. For $W_i \in \mathbb{R}^{d}$, one can leave it as is during training and naturally extract $[0:\mathcal{M}[j]]$ of it during inference.

\subsection{Overall Framework}
The overall framework of \ours is illustrated in \cref{fig:main_structure}, including feature encodings, LRU-based recommendation module, \ours weight masking, etc.
\subsubsection{Language and Image Encoding}
We adopt textural item description as the language input source and image as visual input. Given a metadata dictionary $\mathcal{M}$ containing attributes for each item $i$, we extract its attributes Title, Price, Brand and Categories and perform concatenation of attributes:
\[
\text{Text}_i = \text{Title}_i + \text{Price}_i + \text{Brand}_i + \text{Categories}_i
\]
We then encode these text attributes and image attributes using pretrained embedding models $f$. For each item $i$:
\begin{equation}
    \mathbf{E}_{\text{lang}, i} = f_{\text{lang}}(\text{Text}_i), \; \mathbf{E}_{\text{img}, i} = f_{\text{img}}(\text{Img}_i)
\end{equation}

We combine text and image embedding through concatenation followed by a simple yet effective linear projection:
\begin{equation}
\label{norm combine embedding}
\mathbf{E} = (\text{Concat}(\mathbf{E}_{\text{lang}}, \mathbf{E}_{\text{img}})) \mathbf{W}_{\text{proj}} + \mathbf{b}_{\text{proj}}
\end{equation}
where $\mathbf{W}_{\text{proj}}$ and $\mathbf{b}_{\text{proj}}$ are the projection weights and $\mathbf{W}_{\text{proj}} \in \mathbb{R}^{(D_{\text{lang}} + D_{\text{img}}) \times D}$ and $\mathbf{b}_{\text{proj}} \in \mathbb{R}^{D}$.

\subsubsection{Linear Recurrent Units}
We adopt Linear Recurrent Units (LRU) for sequence processing for its (1) superior performance and (2) both low training and inference cost compared with RNN and Self-Attention-based models~\cite{orvieto2023resurrecting, yue2024linear}. Intuitively, LRU is capable of parallel training like Self-Attention and inference like RNN, where inference complexity can be performed incrementally.

Given input $x_k \in \mathbb{R}^{B \times H_{\mathrm{in}}}$ at time step $k$, hidden state $h_{k-1} \in \mathbb{R}^{B \times H_{\mathrm{in}}}$, learnable matrices $A \in \mathbb{R}^{H \times H_{\mathrm{in}}}$, $B \in \mathbb{R}^{H \times H_{\mathrm{in}}}$, $C \in \mathbb{R}^{H_{\mathrm{out}} \times H_{\mathrm{in}}}$ and $D \in \mathbb{R}^{H_{\mathrm{out}} \times H_{\mathrm{in}}}$:
\begin{equation}
\label{lru sequential}
    \mathbf{h}_{\mathrm{k}} = \mathbf{A} \mathbf{h}_{\mathrm{k-1}} + \mathbf{B} \mathbf{x}_{\mathrm{k}}, \quad \mathbf{y}_{\mathrm{k}} = \mathbf{C} \mathbf{h}_{\mathrm{k}} + \mathbf{D} \mathbf{x}_{\mathrm{k}},
\end{equation}
The input and output dimensions are denoted with $H_{\mathrm{in}}$ and $H_{\mathrm{out}}$ (i.e., embedding size), and the hidden dimension size with $H$. Different from RNN models (i.e., $h_k = \sigma(A h_{k-1} + B x_k)$), we discard the non-linearity $\sigma$ to enable parallelization:
\begin{equation}
\label{lru parallel}
  \begin{aligned}
    \mathbf{h}_{\mathrm{k}} &= \mathbf{A} \mathbf{h}_{\mathrm{k}-1} + \mathbf{B} \mathbf{x}_{\mathrm{k}} \\
        &= \mathbf{A}^2 \mathbf{h}_{\mathrm{k}-2} + \mathbf{A} \mathbf{B} \mathbf{x}_{\mathrm{k}-1} + \mathbf{B} \mathbf{x}_{\mathrm{k}} = \ldots \\
        &= \sum_{\mathrm{i}=1}^{\mathrm{k}} \mathbf{A}^{\mathrm{k}-\mathrm{i}} \mathbf{B} \mathbf{x}_{\mathrm{i}} \quad \mathrm{with} \quad \mathbf{h}_1 = \mathbf{B} \mathbf{x}_1.
  \end{aligned}
\end{equation}
Therefore, LRU can be trained in parallel (via parallel scan) as Self-Attention (equation \ref{lru parallel}) and enable fast inference as RNN models (equation \ref{lru sequential}). 

\subsubsection{Overall LRU-Based Recommendation Framework}
We first pad for the combined embeddings $\mathbf{E}_i$ output by equation \ref{norm combine embedding} to maximum length of all sequences. Then, the padded embeddings $\mathbf{E}_i$ are processed through $N$ blocks. For each block $i \in \{1, \ldots, N\}$, we first perform layer normalization to the input followed by a LRU layer:
\begin{align}
    \text{LayerNorm}(\mathbf{X}) &= \alpha \odot \frac{\mathbf{X}-\mu}{\sqrt{\sigma^2+\epsilon}} + \beta \\
    \text{LRUNorm}(\mathbf{X}) &= \text{LRU}(\text{LayerNorm}(\mathbf{X}))
\end{align}

Due to the lack of non-linearity for LRU, we 
further process the output of LRU layer by a gated non-linear feed-forward network (FFN) to improve training dynamics and model performance. Specifically, our FFN is defined as:
\begin{align*}
\text{Gate} &= \text{SiLU}(\mathbf{X}\mathbf{W}^{(g)} + \mathbf{b}^{(g)})\\
\text{FFN} &= (\text{Gate}\odot(\mathbf{X} \mathbf{W}^{(1)} + \mathbf{b}^{(1)})) \mathbf{W}^{(2)} + \mathbf{b}^{(2)}
\end{align*}
As the network gets deeper, some signal of the input from the earlier layers might be forgotten. Thus, we add sub-layer connections in FFN by adding pre-layer normalization and residual connection:
\begin{equation*}
    \text{SubLayer}(\text{FFN}, \mathbf{X})=\text{FFN}(\text{LayerNorm}(\mathbf{X}))+\mathbf{X}
\end{equation*}

\subsubsection{\ours Plugin to Overall Framework}
Next, we apply fMRLRec-based weight design. Given a set of sizes $\mathcal{M}=\{2,4,8,\ldots,D\}$, any $W_i \in \mathbb{R}_d$, we leave it as is. For $W_i \in \mathbb{R}^{d_1 \times d_2}$, we apply the \ours operator defined in section \ref{sec:fMRLRec} to $W_i$ as:
\begin{equation}
    \mathbf{W'}_i = \text{\ours}(\mathbf{W}_i, \mathcal{M})
\end{equation}

During inference time, \textit{independent} models $\mathcal{Q}=\{\mathcal{W}'^{(1)},\mathcal{W}'^{(2)},\ldots,\mathcal{W}'^{(|\mathcal{M}|)}\}$ could be extracted as described in the last paragraph of section \ref{sec:fMRLRec}.

\paragraph{Prediction Layer} After the final layer $N$, we extract the activation at the last time step $t$ of the final layer as $z_t^{(N)} \in \mathbb{R}^D$, and use it to compute the relevance $r_{i,t} \in \mathbb{R}$ for all items in the pool $v_i \in \mathcal{V}$.
Specifically, we perform dot product between $z_t^{(N)}$ with the input/shared embedding layer weight $E_W \in \mathbb{R}^{|\mathcal{V}| \times D}$:
\begin{equation}
    r_{i,t} = \left(\mathbf{z}_t^{(N)} \mathbf{E}_w^T\right)_i
\end{equation}
The higher $r_{i,t}$, the more likely a user is to consider item $v_i$ for the next time step. This way we could generate recommendations by ranking the relevance score $r_{i,t}$.

\subsubsection{Network Training}
As we derive the relevance score of item $i$ as $r_{i,t}(\theta)$ where $\theta$ stands for all parameters used to compute $r$, we treat the relevance score as logits to compute Cross-Entropy (CE) loss for entire network optimization. 
While LRURec can be trained with CE loss, it is not enough to yield performant models of sizes $\mathcal{M}=\{2,4,8,\ldots,D\}$ as traditional CE loss only explicitly optimizes the largest model of size $D$. We solve this issue by introducing explicit loss terms as introduced in~\cite{kusupati2022matryoshka} to pair with our \ours-style weight matrix for best performance:
\begin{equation}
\mathcal{L}_{\text{\ours}} = \underset{\theta}{\text{min}} \frac{1}{|\mathcal{V}|} \sum_{i = 1}^{|\mathcal{V}|} \sum_{m \in \mathcal{M}} \mathcal{L} \left(\mathbf{r}_{i}(\theta[:m]), \mathbf{y}_i\right)
\end{equation}
where $\mathcal{L}$ is a multi-class softmax cross-entropy loss function based on ranking scores and the label item.

\section{\ours Memory Efficiency}
\label{sec: memory efficiency}
In this subsection, we analyze \ours model-series memory efficiency by driving the number of parameters plus activations needed to train model sizes of $\mathcal{M}=\{2,4,8,\ldots,D\}$ or $\mathcal{M}=\{2^j|j=1,2,\ldots,k\}$ as (1) A train-once \ours model-series and (2) Independent models. Define $W^{(j)}=\{w_1^{(j)},w_2^{(j)},\cdot,w_n^{(j)}\}$ as the layer weights of model size $j$ and $X_i \in \mathbb{R}^{B \times L \times D}$ as sequential input data for $w_i$, where $B$ is the batch size, $L$ is the sequence length and $D=2^j$ is the model size. We assume every weight has the same scaling factor $\gamma$ to simplify notations. Thus, $\gamma \cdot (2^j)^2$ and $\gamma \cdot 2^j$ are number of parameters for 2d and 1d weight. Here, we only consider 2d weights saves the most parameters.

Case 1: For \ours-based training, number of parameters needed $N(W)=\sum_{i=1}^n \gamma (\cdot (2^k)^2)$, which is $n \cdot \gamma \cdot 2^{(2k)}$; The number of activation generated $N(A) = \sum_{i=1}^n \gamma \cdot B \cdot L \cdot D$. Empirically, $B \in \{32, 64, 128\}$ and $L=50$, thus $B \cdot L = \delta \cdot 2^k$, $\delta>1$. Then, we have $N(A) = n \cdot \gamma \cdot \delta \cdot 2^{(2k)}$.

Case 2: For Independent training, the number of parameters needed $N(W)=\sum_{j=1}^k \sum_{i=1}^n \gamma \cdot (2^j)^2$, by summation of the geometric series, $N(W)=n \cdot \gamma \cdot \frac{4^{k+1}-4}{3}$, the number of activation generated $N(A) = \sum_{j=1}^k \sum_{i=1}^n \gamma \cdot B \cdot L \cdot D$.  Empirically, $B \in \{32, 64, 128\}$ and $L=50$, thus $B \cdot L = \delta \cdot 2^j$, $\delta>1$. Then, we have $N(A) = \sum_{j=1}^k n \cdot \gamma \cdot \delta \cdot 2^{(2j)}$, which is equivalent to $N(A) = n \cdot \gamma \cdot \delta \cdot \frac{4^{k+1}-4}{3}$.

In summary, the ratio of parameters and activations between \ours-based training and Independent training is $R=(n \cdot \gamma \cdot \frac{4^{k+1}-4}{3}) / (n \cdot \gamma \cdot 2^{(2k)})$ or $(n \cdot \gamma \cdot \delta \cdot \frac{4^{k+1}-4}{3}) / (n \cdot \gamma \cdot \delta \cdot 2^{(2k)}) \approx 1.33$. This indicates a parameter saving rate $R_s$ of $\approx 0.33$ against the \ours model. Empirically, for a common setting $n=4$ linear layers with scaling factor $\gamma=2$ and $D=512$, the weights saved are approximately $4(n) \times 0.33(R) \times 512(D) \times 1024(2D) \approx 700K$, the number of activation saved for four layer is approximately $ 4(n) \times 0.33(R) \times 32(B) \times 50(L) \times 1024(2D) \approx 2M$. This is to a great extent saving memory usage if independent training is executed in parallel or saving training time if executed sequentially.
\section{Experimental Setup}
\label{sec:exp setup}
\begin{table}[t]
\centering
\caption{Statistics of the datasets.}
\resizebox{1.0\linewidth}{!}{
\begin{tabular}{@{}lccccc@{}}
\toprule
\textbf{Name}     & \textbf{\#User} & \textbf{\#Item} & \textbf{\#Image} & \textbf{\#Inter.} & \textbf{Density} \\ \midrule
\textbf{Beauty}   & 22,363          & 12,101          & 12,023           & 198k              & 0.073           \\
\textbf{Clothing} & 39,387          & 23,033          & 22,299           & 278k              & 0.031           \\
\textbf{Sports}   & 35,598          & 18,357          & 17,943           & 296k              & 0.045           \\
\textbf{Toys}     & 19,412          & 11,924          & 11,895           & 167k              & 0.072           \\ \bottomrule
\end{tabular}
}
\label{tab:data_stat}
\end{table}
\begin{table*}[t]
\centering
\caption{Main performance results of \ours and baselines.}
\resizebox{\textwidth}{!}{%
\begin{tabular}{llcccccccccc}
\toprule
\multirow{2}{*}{\textbf{Dataset}} & \multirow{2}{*}{\textbf{Metric}} & \multicolumn{4}{c}{\textbf{ID-Based}}   & \multicolumn{3}{c}{\textbf{Text-Based}} & \multicolumn{3}{c}{\textbf{Multimodal}} \\ \cmidrule(l){3-6} \cmidrule(l){7-9} \cmidrule(l){10-12}
                                  &                                  & SAS    & BERT   & FMLP   & LRU          & UniS.     & VQRec     & RecF.           & MMSSL  & VIP5         & fMRLRec         \\ \midrule
\textbf{Beauty}                   & N@5                              & 0.0274 & 0.0275 & 0.0318 & 0.0339          & 0.0274  & 0.0303        & 0.0258        & 0.0189 & \ul{0.0339} & \textbf{0.0415} \\
                                  & R@5                              & 0.0456 & 0.0420 & 0.0539 & \ul{0.0565}    & 0.0484  & 0.0514        & 0.0428        & 0.0308 & 0.0417       & \textbf{0.0613} \\
                                  & N@10                             & 0.0364 & 0.0350 & 0.0416 & \ul{0.0438}    & 0.0375  & 0.0411        & 0.0341        & 0.0252 & 0.0367       & \textbf{0.0520} \\
                                  & R@10                             & 0.0734 & 0.0653 & 0.0846 & \ul{0.0871}    & 0.0799  & 0.0849        & 0.0686        & 0.0506 & 0.0603       & \textbf{0.0939} \\ \midrule
\textbf{Cloth.}                   & N@5                              & 0.0075 & 0.0062 & 0.0091 & 0.0104          & 0.0127  & 0.0104        & \ul{0.0137}  & 0.0089 & 0.0122       & \textbf{0.0193} \\
                                  & R@5                              & 0.0134 & 0.0100 & 0.0167 & 0.0192          & 0.0221  & 0.0197        & \ul{0.0234}  & 0.0146 & 0.0152       & \textbf{0.0333} \\
                                  & N@10                             & 0.0104 & 0.0084 & 0.0123 & 0.0140          & 0.0175  & 0.0149        & \ul{0.0192}  & 0.0122 & 0.0183       & \textbf{0.0259} \\
                                  & R@10                             & 0.0227 & 0.0169 & 0.0266 & 0.0304          & 0.0372  & 0.0336        & \ul{0.0405}  & 0.0249 & 0.0298       & \textbf{0.0541} \\ \midrule
\textbf{Sports}                   & N@5                              & 0.0143 & 0.0137 & 0.0194 & \ul{0.0204}    & 0.0141  & 0.0173        & 0.0127        & 0.0123 & 0.0136       & \textbf{0.0230} \\
                                  & R@5                              & 0.0267 & 0.0215 & 0.0329 & \ul{0.0344}    & 0.0237  & 0.0304        & 0.0211        & 0.0198 & 0.0264       & \textbf{0.0349} \\
                                  & N@10                             & 0.0210 & 0.0181 & 0.0252 & \ul{0.0266}    & 0.0195  & 0.0235        & 0.0173        & 0.0163 & 0.0213       & \textbf{0.0284} \\
                                  & R@10                             & 0.0474 & 0.0355 & 0.0508 & \textbf{0.0536} & 0.0408  & 0.0497        & 0.0350        & 0.0321 & 0.0315       & \ul{0.0516}    \\ \midrule
\textbf{Toys}                     & N@5                              & 0.0291 & 0.0241 & 0.0308 & \ul{0.0366}    & 0.0254  & 0.0314        & 0.0292        & 0.0173 & 0.0334       & \textbf{0.0461} \\
                                  & R@5                              & 0.0534 & 0.0355 & 0.0534 & \ul{0.0601}    & 0.0477  & 0.0577        & 0.0501        & 0.0286 & 0.0474       & \textbf{0.0672} \\
                                  & N@10                             & 0.0380 & 0.0299 & 0.0408 & \ul{0.0463}    & 0.0362  & 0.0423        & 0.0398        & 0.0224 & 0.0374       & \textbf{0.0552} \\
                                  & R@10                             & 0.0807 & 0.0535 & 0.0845 & 0.0901          & 0.0811  & \ul{0.0915}  & 0.0832        & 0.0445 & 0.0642       & \textbf{0.0956} \\ \midrule
\textbf{Avg.}                     & N@5                              & 0.0196 & 0.0179 & 0.0228 & \ul{0.0253}    & 0.0199  & 0.0224        & 0.0204        & 0.0144 & 0.0233       & \textbf{0.0325} \\
                                  & R@5                              & 0.0348 & 0.0273 & 0.0392 & \ul{0.0426}    & 0.0355  & 0.0398        & 0.0344        & 0.0235 & 0.0327       & \textbf{0.0492} \\
                                  & N@10                             & 0.0265 & 0.0229 & 0.0300 & \ul{0.0327}    & 0.0277  & 0.0305        & 0.0276        & 0.0191 & 0.0284       & \textbf{0.0404} \\
                                  & R@10                             & 0.0561 & 0.0428 & 0.0616 & \ul{0.0653}    & 0.0598  & 0.0649        & 0.0568        & 0.0381 & 0.0465       & \textbf{0.0738} \\ \bottomrule
\end{tabular}%
}
\label{tab:main results}
\end{table*}
\subsection{Datasets}
For evaluating our models, we select four commonly used benchmarks from \emph{Amazon.com} known for real-word sparsity, namely \emph{Beauty}, \emph{Clothing, Shoes \& Jewelry} (Clothing), \emph{Sports \& Outdoors} (Sports) and \emph{Toys \& Games} (Toys)~\cite{mcauley2015image, he2016ups}. For preprocessing, we follow~\cite{yue2022defending, chen2023palr, geng2023vip5} to construct the input sequence in chronological order and apply 5-core filtering to exclude users and items with less than five-time appearances. For textural feature selection, we choose \emph{title}, \emph{price}, \emph{brand} and \emph{categories}; For visual features, we use \emph{photos} of the items. We also filter out items without above metadata. Detailed statistics of the datasets are reported in table \ref{tab:data_stat} including users (\#User), items (\#Item), images (\#Image), interactions (\#Inter.) and dataset density in percentages.

\subsection{Baseline Methods}
For baseline models, we select a series of state-of-the-art recommendation models grouped as \emph{ID-based}, \emph{Text-based} and \emph{Multimodal}. \emph{ID-based} models include SASRec, BERT4Rec, FMLP-Rec and LRURec~\cite{kang2018self, sun2019bert4rec, zhou2022filter, yue2024linear}. Text-based methods include UniSRec, VQRec and RecFormer~\cite{hou2022towards, hou2023learning, li2023text}. We also include multimodal baselines MMSSL, VIP5~\cite{wei2023multi, geng2023vip5}, More details about baselines is discussed in \Cref{sec:append:baselines}.

\subsection{Implementations}
For training \ours and all baseline models, we utilize AdamW optimizer with learning rate of 1e-3/1e-4 with maximum epochs of 500. Validation is performed per epoch and the training is stopped once validation performance does not improve for 10 epochs. The model with best validation performance is saved for testing and metrics report. For hyperparameters, we find (1) embedding/model size, the number of \ours-LRU layers, dropout rate and weight decay be the most sensitive ones for model performance. Specifically, we grid-search the embedding/model size in [64, 128, 256, 512, 1024, 2048], the number of \ours-LRU layers in [1,2,4,8], dropout rate from [0.1,...,0.8] on a 0.1-stride and weight decay from [1e-6, 1e-4, 1e-2]. For ring-initialization of LRU layers, we grid-search the minimum radius in [0.0,...,0.5] on a 0.1-stride. The max radius is set to the minimum radius plus 0.1. The best hyper-parameters for each datasets are reported in \Cref{sec:append:implementations}; We follow~\cite{geng2023vip5} and set maximum length of input sequence as 50. For validation and test, we adopt two metrics NDCG@$k$ and Recall@$k$, $k \in \{5,10\}$ typical for recommendation algorithm evaluation.
\section{Experimental Results}
\subsection{Main Performance Analysis}
Here, we compare the performance of \ours with state-of-the-art baseline models in table \ref{tab:main results}. We use SAS, BERT, FMLP, LRU, UniS., RecF., \ours to abbreviate SASRec BERT4Rec, FMLP-Rec LRURec, UniSRec, RecFormer and \ours. The best metrics are marked in bold and the second best metrics are underlined. Overall, \ours outperforms all baseline models in almost all cases with exceptions of Recall@10 for Sports. Specifically, We observe that: (1) \ours on average outperforms the second-best model by 17.98\% across all datasets and metrics (2) \ours shows superior ranking performance by having a more significant gain of NDCG which is ranking sensitive than Recall. For example, \ours achieves NDCG@5 improvement of 25.42\% over the second best model, which is greater than the Recall@5 gains of 16.01\%. This is also true for NDCG@10 gains of 19.97\% compared with recall gains of 10.51\%. (3) \ours demonstrates significant benefits for sparse datasets, Clothing and Sports, by averaging 21.11\% improvements. In contrast, the average gains is lower as 14.84\% for relatively denser datasets as Beauty and Toys. In summary, our results suggest \ours can effectively leverage multimodal item representation to rank items of user preference and improve recommendation performance.
\begin{figure}[t]
    \centering
    \begin{subfigure}[]{0.49\linewidth}
        \centering
        \includegraphics[width=\textwidth]{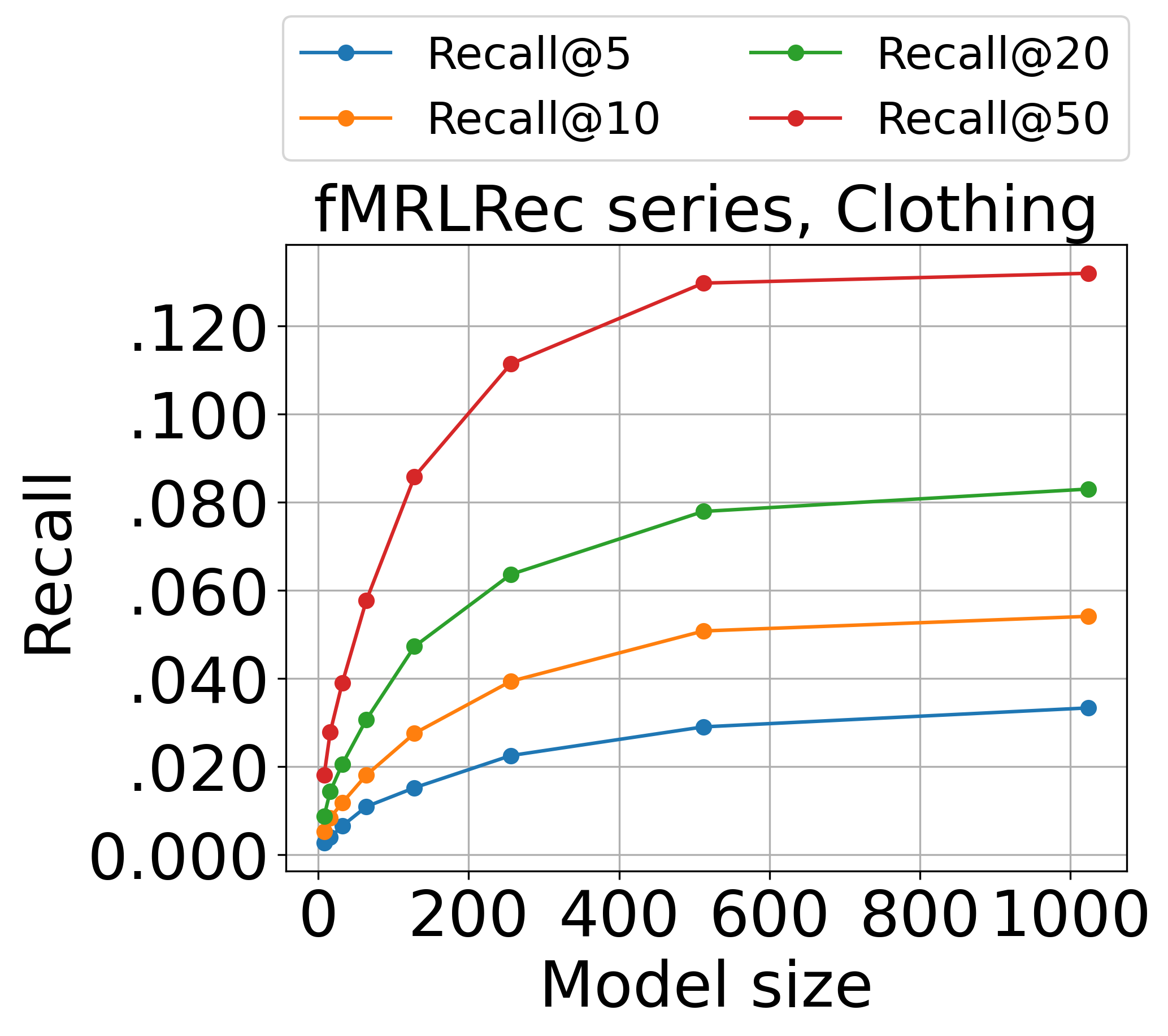}
        \caption{Recall for Clothing}
        \label{fig:recall clothing}
    \end{subfigure}
    \begin{subfigure}[]{0.49\linewidth}
        \centering
        \includegraphics[width=\textwidth]{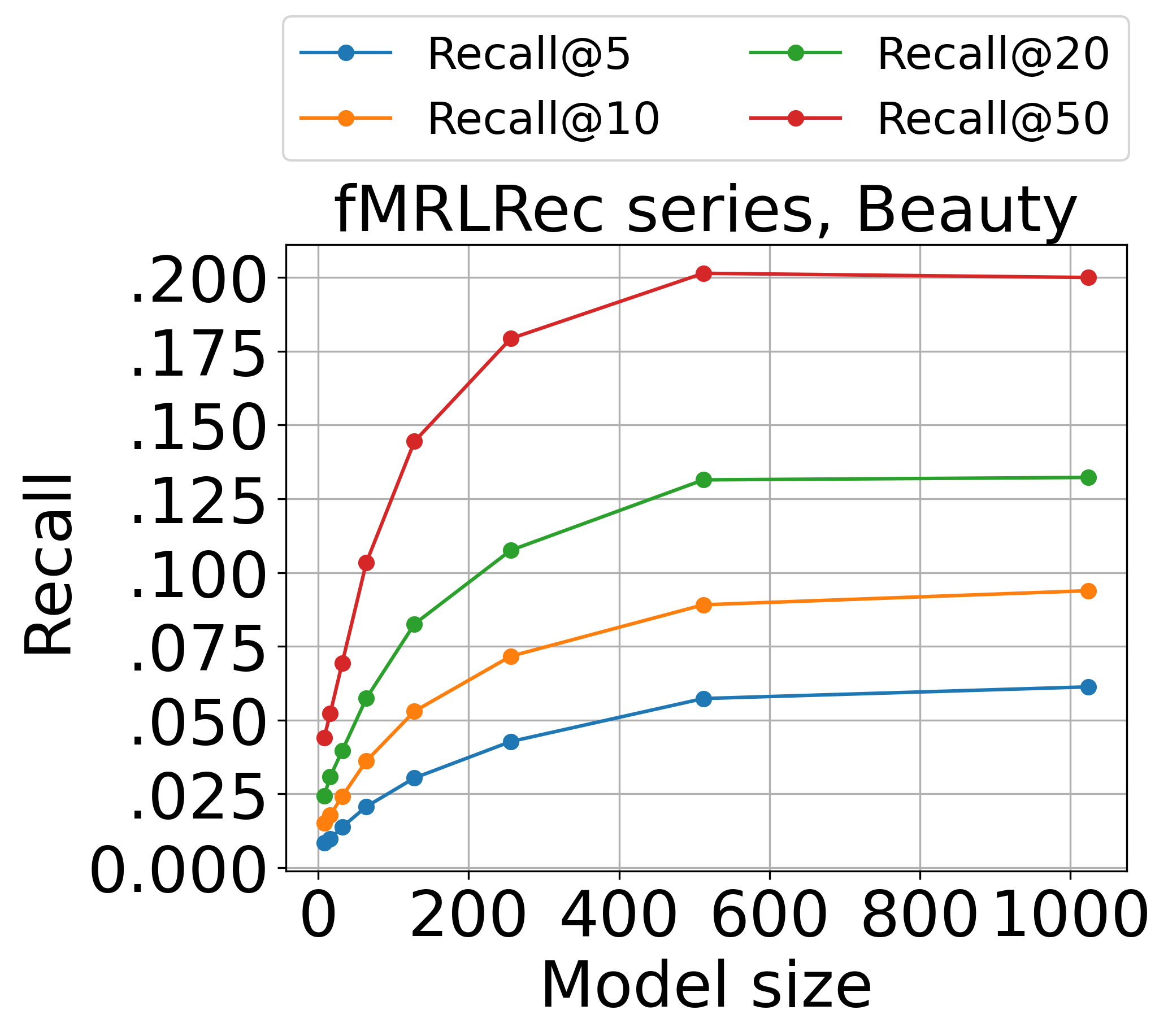}
        \caption{Recall for Beauty}
        \label{fig:recall beauty}
    \end{subfigure}
    \begin{subfigure}[]{0.49\linewidth}
        \centering
        \includegraphics[width=\textwidth]{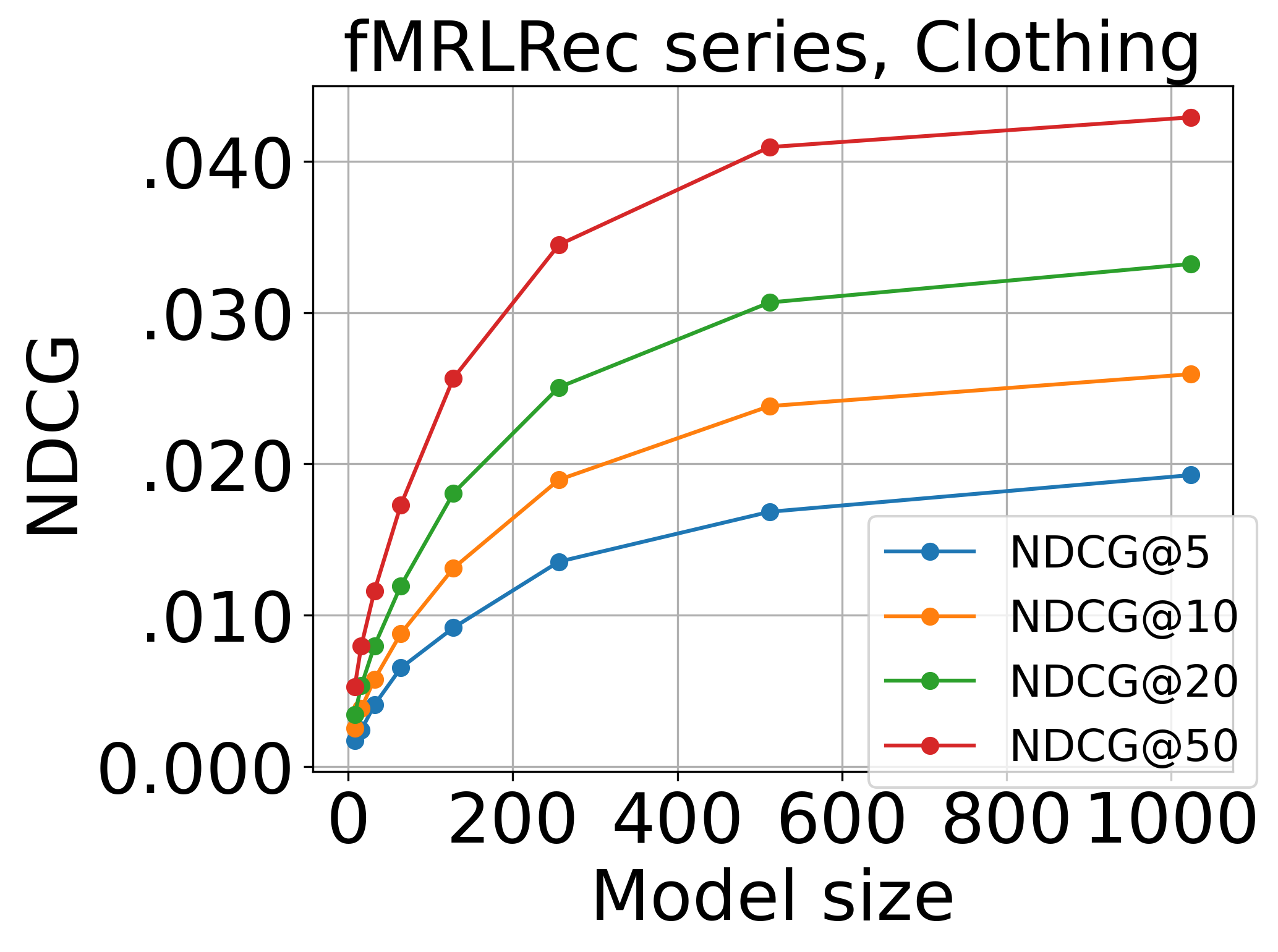}
        \caption{NDCG for Clothing}
        \label{fig:ndcg clothing}
    \end{subfigure}
    \begin{subfigure}[]{0.49\linewidth}
        \centering
        \includegraphics[width=\textwidth]{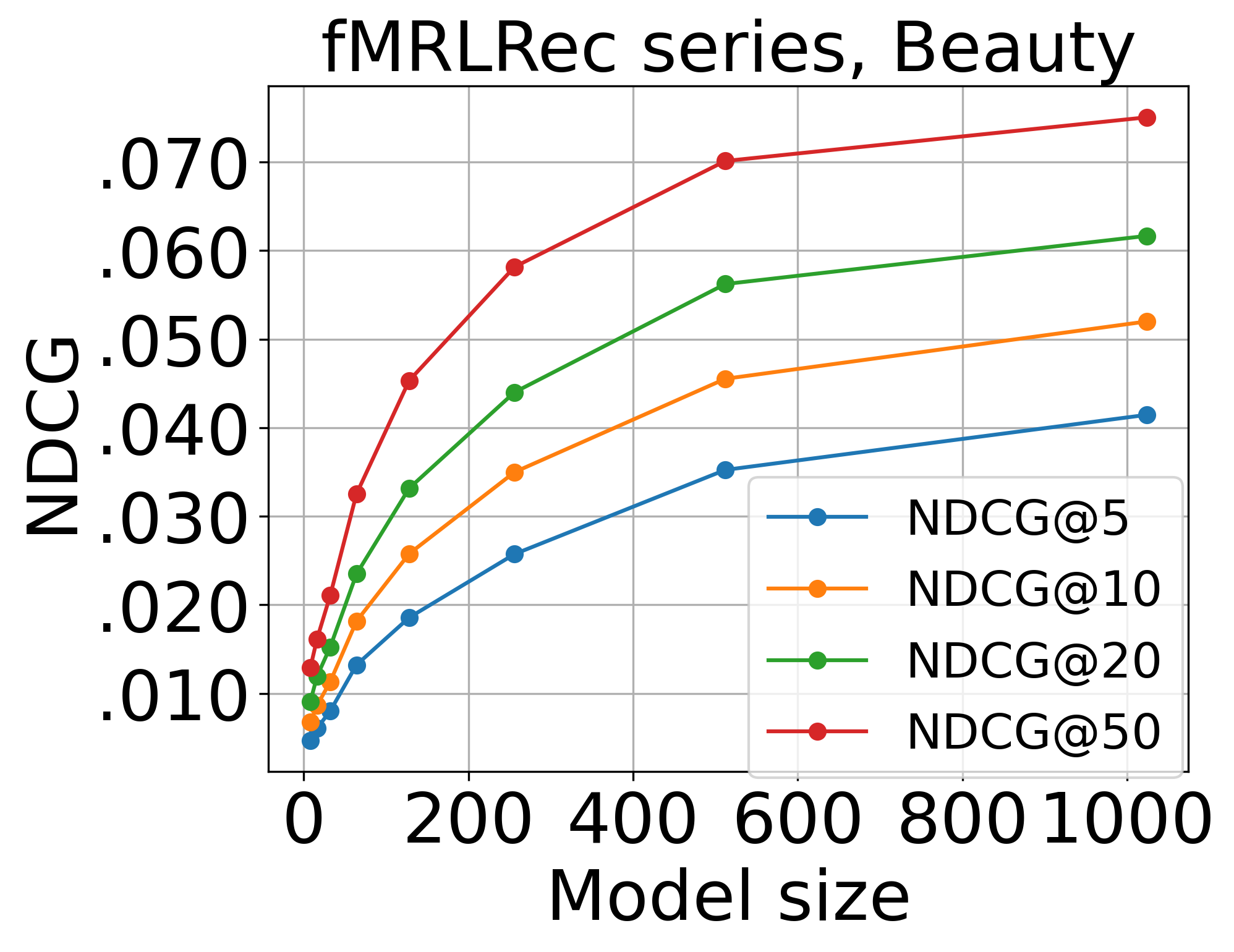}
        \caption{NDCG for Beauty}
        \label{fig:ndcg beauty}
    \end{subfigure}
    \caption{fMRLRec-model series performance curve against model size. fMRLRec features a significantly slower performance drop for example with drop rates from 6.14\% to 37.69\% (Recall@10 for Clothing) compared to the model compression rate of 50\%.}
\label{fig:fMRL curve}
\end{figure}

\subsection{\ours Model-Series Performance}
\begin{table*}[t]
\caption{Ablation performance for fMRLRec by removing either of language (lang.) or visual features or both.}
\centering
\resizebox{\textwidth}{!}{%
\begin{tabular}{llcccccccc}
\toprule
\multicolumn{2}{c}{\textbf{Variants / Dataset}}            & \multicolumn{2}{c}{\textbf{Beauty}} & \multicolumn{2}{c}{\textbf{Clothing}} & \multicolumn{2}{c}{\textbf{Sports}} & \multicolumn{2}{c}{\textbf{Toys}} \\ \cmidrule(lr){1-2} \cmidrule(lr){3-4} \cmidrule(lr){5-6} \cmidrule(lr){7-8} \cmidrule(l){9-10}
\multicolumn{2}{c}{\textbf{Metric}}                        & NDCG             & Recall           & NDCG              & Recall            & NDCG             & Recall           & NDCG            & Recall          \\ \midrule
\multirow{2}{*}{\textbf{fMRLRec}}                    & @5  & \textbf{0.0415}  & \textbf{0.0613}  & \textbf{0.0193}   & \textbf{0.0333}   & \textbf{0.0230}  & \textbf{0.0349}  & \textbf{0.0461} & \textbf{0.0672} \\
                                                     & @10 & \textbf{0.0520}  & \textbf{0.0939}  & \textbf{0.0259}   & \textbf{0.0541}   & \textbf{0.0284}  & \textbf{0.0516}  & \textbf{0.0552} & \textbf{0.0956} \\ \midrule
\multirow{2}{*}{\textbf{fMRLRec w/ Lang. only}}      & @5  & 0.0353           & \ul{ 0.0561}     & \ul{ 0.0167}      & \ul{ 0.0279}      & \ul{ 0.0205}     & \ul{ 0.0313}     & 0.0403          & \ul{ 0.0618}    \\
                                                     & @10 & 0.0449           & \ul{ 0.0859}     & \ul{ 0.0225}      & 0.0461            & \ul{ 0.0261}     & \ul{ 0.0487}     & 0.0503          & \ul{ 0.0927}    \\ \midrule
\multirow{2}{*}{\textbf{fMRLRec w/ Image only}}      & @5  & \ul{ 0.0370}     & 0.0540           & 0.0162            & 0.0279            & 0.0194           & 0.0291           & \ul{ 0.0416}    & 0.0613          \\
                                                     & @10 & \ul{ 0.0464}     & 0.0833           & 0.0222            & \ul{ 0.0467}      & 0.0238           & 0.0430           & \ul{ 0.0516}    & 0.0920          \\ \midrule
\multirow{2}{*}{\textbf{fMRLRec w/o Lang. \& Image}} & @5  & 0.0257           & 0.0335           & 0.0035            & 0.0046            & 0.0113           & 0.0153           & 0.0287          & 0.0350          \\
                                                     & @10 & 0.0288           & 0.0431           & 0.0040            & 0.0062            & 0.0127           & 0.0197           & 0.0309          & 0.0418          \\ \bottomrule
\end{tabular}%
}
\label{tab:ablation results}
\end{table*}
In this subsection, we analyze the performance of our full scale Matryoshka Representation Learning (\ours) by extracting from trained models the differently-sized sub-models of $\mathcal{M}=\{8,16,32,\ldots,D\}$, where $D=1024$ here for best performance. Specific sub-model performance is shown in figure \ref{fig:fMRL curve}. Using Recall for Clothing as an example, we observe that: (1) The Recall decrease rate for Clothing ranges from 6.14\% to 37.69\% which is significantly lower than the exponential model compressed by a rate of 50\%. This is consistent with the \textit{Scaling Law}~\cite{kaplan2020scaling} that doubling the model size usually does not mean doubling performance. Despite statement of the \textit{Scaling Law}, the specific performance retained varies for datasets/tasks and are expensive to tune. Tackling this pain point, \ours curve in figure \ref{fig:fMRL curve} provides flexible options of how much metric score to retain for developers with limited computational resources. And obtaining \ours such patterns only requires a one-time training of the largest model as introduced in section \ref{sec:fMRLRec}.

\subsection{Parameter Saving of \ours}
\begin{figure}[t]
    \centering
    \includegraphics[width=0.7\linewidth]{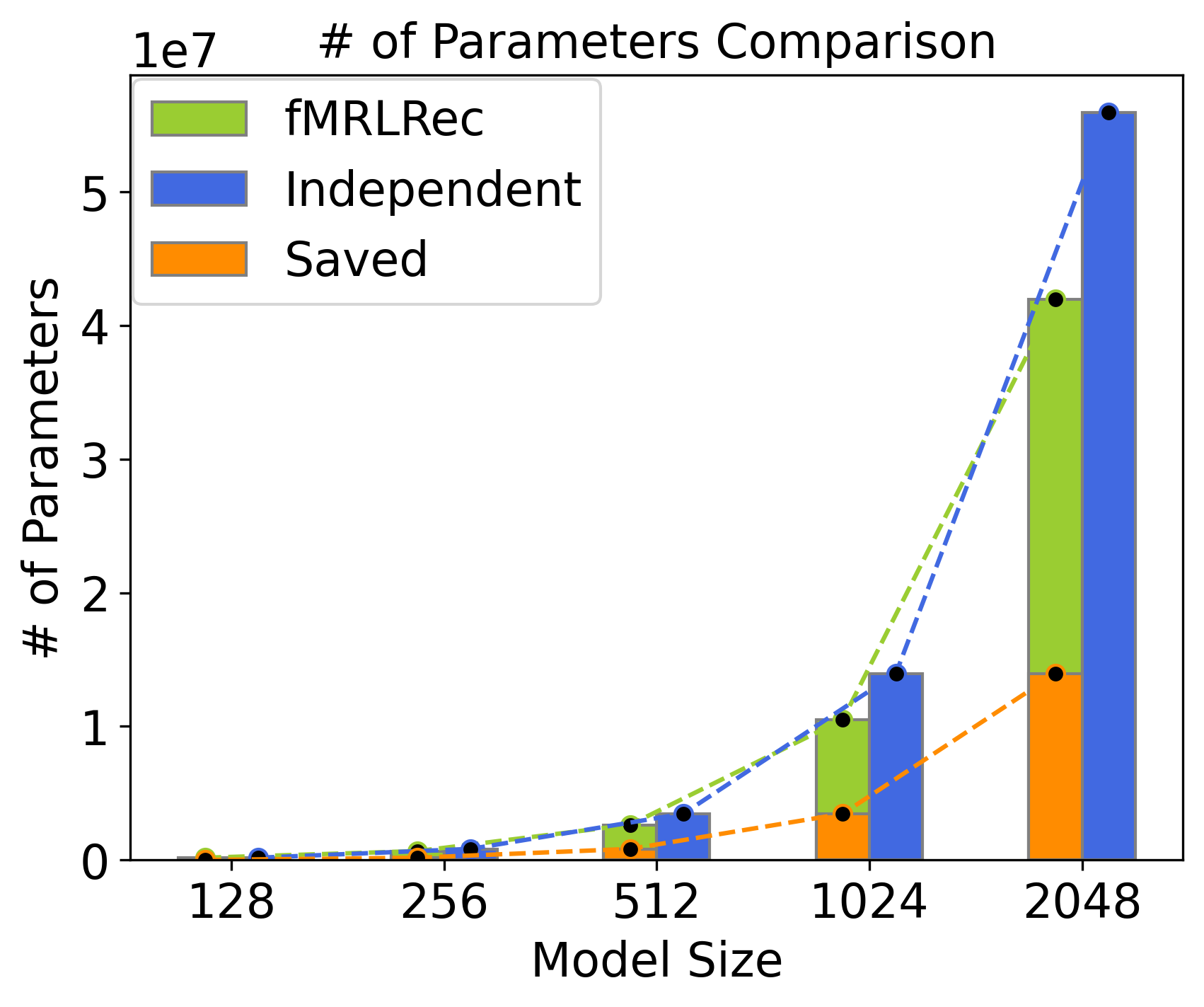}
    \caption{fMRL features a one-time training of model sizes $\mathcal{M}=\{2,4,\ldots,2^n\}$ that saves $\approx33\%$ parameters compared to training every size independently.} 
    \label{fig:num model params}
\end{figure}

Discussed in \Cref{sec: memory efficiency}, the model parameter saving rate $R_s$ between \ours-model series and independently trained models is theoretically around $1/3$ of the former. We demonstrate in figure \ref{fig:num model params} this behavior given model sizes of $\mathcal{M}=\{2^7, 2^8, \ldots, 2^{11}\}$. The green, blue and orange bar represents the number of parameters of \ours-series, independently trained models and ones saved, respectively. Empirically, $R_s = [0, 25.16\%, 31.39\%, 32.90\%, 33.25\%]$ for $\mathcal{M}[j] \in \mathcal{M}$, which converges to $\approx 0.33$ as $j$ gets larger and is consistent with our theoretical analysis in \Cref{sec: memory efficiency}.
\subsection{Ablation Study}
In this section, we further evaluate the designs of features and modules of \ours by a series of ablation studies in \cref{tab:ablation results}. Specifically, we construct different variants of \ours as: (1) \ours w/ Language only: the \ours model with only the text-based attributes of items such as Title, brand, etc. and their corresponding embeddings. (2) \ours w/ Image only: the \ours model only with the image processor and embeddings. (3) \ours w/o Language \& Image: \ours removing all the language and image related feature processing and embeddings. A randomly initialized embedding table is used as item representations. We monitor the change of NDCG and Recall of above variants. In particular, (1) Language features show a predominant contribution for the overall performance as removing language features (\ours w/ Image only) induces the largest performance drop of 12.45\%. (2) Image feature also constitute a vital but relatively lighter contribution compared with language features with a performance drop of 10.67\% when removed (\ours w/ Lang. only); (3) Losing both image and language features induces the largest performance drop of 58.35\% which justifies contributions of both modalities; In summary, our ablation results show that both language and image feature processing and fusion are effective towards improving the recommendation performance of \ours.

\section{Conclusions}
In this work, we introduce a lightweight framework \ours for efficient multimodal recommendation across multiple granularities. In particular, we adopt Matryoshka representation learning and design an efficient linear transformation to embeds smaller features into larger ones. Moreover, we incorporate cross-modal features and further improves the state-space modeling for sequential recommendation. Consequently, \ours can yield multiple model sizes with competitive performance within a single training session. To validate the effectiveness and efficiency of \ours, we conducted extensive experiments, where \ours consistently demonstrate the superior performance over state-of-the-art baseline models.

\section{Limitations}
We have discussed the the ability of \ours to perform one-time training and yield models in multiples sizes ready for deployment. However, we have not experimented on other recommendation tasks such as click rate prediction and multi-basket recommendation, etc. Even though we adopted LRU, a state-of-the-art recommendation module for \ours, other types of sequential/non-sequential models needs to be tested for a more compete performance pattern. More broadly, The idea of full-Scale Matryoshka Representation Learning (fMRL) can be applied to other ML domains that utilize neural network weights; We have yet to explore behaviors of fMRL in those fields where the scale of models and data varies significantly. We plan to conduct more theoretical analysis and experiments for above mentioned aspects in future works.

\bibliography{custom}

\appendix
\section{Appendix}
\label{sec:appendix}

\subsection{Baselines}
\label{sec:append:baselines}
We select multiple state-of-the-art baselines to compare with \ours. In particular, we adopt \textit{ID-based} SASRec, BERT4Rec, FMLP-Rec and LRURec~\cite{kang2018self, sun2019bert4rec, zhou2022filter, yue2024linear}, text-based UniSRec, VQRec and RecFormer~\cite{hou2022towards, hou2023learning, li2023text}, and multimodal baselines MMSSL, VIP5~\cite{wei2023multi, geng2023vip5}. We report the details of baseline methods:
\begin{itemize}[leftmargin=10pt]
    \item \emph{Self-Attentive Sequential Recommendation (SASRec)} is the first transformer-based sequential recommender. SASRec uses unidirectional self-attention to capture transition patterns~\cite{kang2018self}.
    \item \emph{Bidirectional Encoder Representations from Transformers for Sequential Recommendation (BERT4Rec)} is similar to SASRec but utilizes bidirectional self-attention. BERT4Rec learns via masked training~\cite{sun2019bert4rec}.
    \item \emph{Filter-enhanced MLP for Recommendation (FMLP-Rec)} also adopts an all-MLP architecture with filter-enhanced layers. FMLP-Rec also applies Fast Fourier Transform (FFT) to improve representation learning~\cite{zhou2022filter}.
    \item \emph{Linear Recurrence Units for Sequential Recommendation (LRURec)} is based on linear recurrence and is optimized for parallelized training. LRURec thus provides both efficient training and inference speed~\cite{yue2024linear}.
    \item \emph{Universal Sequence Representation for Recommender Systems (UniSRec)} is a text-based recommender system. UniSRec leverage pretrained language models to generate item features for next-item prediction~\cite{hou2022towards}.
    \item \emph{Vector-Quantized Item Representation for Sequential Recommenders (VQRec)} is also text-based sequential recommender. VQRec quantizes language model-based item features to improve performance~\cite{hou2023learning}.
    \item \emph{Language Representations for Sequential Recommendation (RecFormer)} is language model-based architecture for recommendation. RecFormer adopts contrastive learning to improve item representation~\cite{li2023text}.
    \item \emph{Multi-Modal Self-Supervised Learning for Recommendation (MMSSL)} is a multimodal recommender using graphs and multimodal item features for recommendation. MMSSL is trained in a self-supervised fashion~\cite{wei2023multi}.
    \item \emph{Multimodal Foundation Models for Recommendation (VIP5)} is a multimodal recommender using item IDs and multimodal attributes for multi-taks recommendation. VIP5 is trained via conditional generation~\cite{geng2023vip5}.
\end{itemize}
All models are trained according to the methodologies described in the original works, with unspecified hyperparameters used as recommended. All baseline methods and \ours are evaluated under identical conditions. 

\subsection{Implementations}
\label{sec:append:implementations}
We discuss further implementation details other than data processing, evaluation metrics, early stopping, etc., as already reported in section \ref{sec:exp setup}. We adopt pretrained BAAI/bge-large-en-v1.5~\cite{xiao2024c} and SigLip~\cite{zhai2023sigmoid} for language and image encoding; The tuning phase basically lasts for 5-6 hours on a single NVIDIA-A100 (40GB) GPU.  For hyperparameters, we find the most sensitive ones towards performance as follows and report the best hyper-parameters found: 
\begin{itemize}
    \item Embedding/model size: We grid-search Embedding/model size among [64, 128, 256, 512, 1024, 2048], the best performing values is 1024 for all datasets, namely Beauty, Clothing, Sport and Toys. This shows that our \ours scales well to large dimensions of pretrained vision/language models with effective modality alignment.
    \item The number of \ours-based LRU layers: We grid-search the number of layers among [1,2,4,8]. The best performing value is 2 for all datasets.
    \item Dropout rate: We grid-search the dropout rate among [0.1,0,2, ..., 0.8] on a 0.1-stride. We find dropout rates, 0.5 or 0.6, is typically optimal for all datasets. 
    \item Weight decay: We grid-search the weight decay among [1e-6, 1e-4, 1e-2] and finds 1e-2 to be the best performing value.

    \item Radius of ring-initialization: For ring initialization of LRU layers, We grid-search the minimum radius of the ring in [0.0,...,0.5] on a 0.1-stride and set the maximum radius to the minimum radius plus 0.1. The best minimum radius is 0.0, 0.1, 0.1, 0.0 for Beauty, Clothing, Sports, Toys, respectively.
\end{itemize}

\end{document}